\documentclass[fleqn,usenatbib]{mnras}

\usepackage[T1]{fontenc}

\DeclareRobustCommand{\VAN}[3]{#2}
\let\VANthebibliography\thebibliography
\def\thebibliography{\DeclareRobustCommand{\VAN}[3]{##3}\VANthebibliography}

\usepackage{graphicx}	
\usepackage{amsmath}	
\usepackage{amssymb}	
\usepackage{newtxtext,newtxmath}

\title[Duration-energy relation of CHIME FRBs]{On the relation between duration and energy of non-repeating fast radio bursts: census with the CHIME data}
    
\author[S. J. Kim et al.]{
Seong Jin Kim$^{1}$\thanks{E-mail: seongini@gmail.com},
Tetsuya Hashimoto$^{2}$,
Bo Han Chen$^{1}$,
Tomotsugu Goto$^{1}$,
Simon C.-C. Ho$^{1}$,
\newauthor
Tiger Yu-Yang Hsiao$^{1}$,
Yi Hang Valerie Wong$^{1}$,
Shotaro Yamasaki$^{2}$
\\
$^{1}$Institute of Astronomy, National Tsing Hua University, 101, Section 2. Kuang-Fu Road, Hsinchu, 30013, Taiwan (R.O.C.)\\
$^{2}$Department of Physics, National Chung Hsing University, 145 Xingda Rd., South Dist., Taichung 40227, Taiwan\\
}

\date{Accepted 2022 June 09. Received 2022 June 09; in original form 2021 November 16 }

\pubyear{2022}

\begin{document}
\label{firstpage}
\pagerange{\pageref{firstpage}--\pageref{lastpage}}
\maketitle

\begin{abstract}
A correlation between the intrinsic energy and the burst duration of non-repeating fast radio bursts (FRBs) has been reported.  If it exists, the correlation can be used to estimate  intrinsic energy from the duration, and thus can provide us with a new distance measure for cosmology. However, the correlation suffered from small number statistics (68 FRBs) and was not free from contamination by latent repeating populations, which might not have such a correlation.  How to separate/exclude the repeating bursts from the mixture of all different types of FRBs   is essential to see this property.
Using a much larger sample from the new FRB catalogue (containing 536 FRBs) recently released by the CHIME/FRB Project, combined with a new classification method  developed based on unsupervised machine learning, we carried out further scrutiny of the relation.   We found that there is a weak correlation between the intrinsic energy and duration for non-repeating FRBs at $z < 0.3$ with Kendall's $\tau$ correlation coefficient of $0.239$ and significance  $0.001$ (statistically significant), whose slope looks similar to that of gamma-ray bursts.  This correlation becomes weaker and insignificant at higher redshifts ($z > 0.3$),  possibly due  to the lack of the faint FRBs at high-$z$ and/or the redshift evolution of the correlation.   The `scattering time' in the CHIME/FRB catalogue shows an intriguing trend: it varies along the line obtained from linear fit on the energy versus duration plane between these two parameters.  A possible cosmological application of the relation must wait for faint FRBs at high-$z$.
\end{abstract}




\section{Introduction}
\label{introduction}

Fast Radio Bursts \citep[FRBs;][]{Lorimer2007},   $\sim$ millisecond (ms) duration radio flashes, are a newly emerging population since they were discovered over a decade ago. Their origin has not been completely revealed and is still a mystery. Despite a large number of physical models proposed for FRBs \citep[e.g.,][]  {Platts2019, Zhang2020}, none of them is satisfactory because they cannot explain all the observed characteristics. This may imply that FRBs could be classified into a number of different types.

Astronomers have already been using FRBs to investigate the materials through which the signal have travelled, which could revolutionize studies of the baryonic matter in the Universe \citep[e.g.,][]{Macquart2020, Hashimoto2020b}.  These efforts are currently extending to researches on the early stage of the Universe, suggesting FRBs as a new tool for distance measure,   giving us $H_0$ measurements, or even constraining the dark energy of the Universe \citep[]{Hashimoto2019, Hashimoto2021, WuQ2020, Hagstotz2021}.
Because  more distant FRBs should have  larger dispersion measures (DM) due to the larger amount of intergalactic medium between the FRBs and us, the observed DM of FRBs can be an alternative and useful indicator of distance to FRBs.  

Any observed relation (if it exists) between the physical parameters of the current FRB samples would be interesting because it might bring us a clue to a better way to describe their characteristics, where our extended understanding could lead to a powerful approach to reveal the physics and origins cloaked behind the phenomena.  
A relation between the radio energy and the duration ($w$) recently reported by \citet[]{Hashimoto2019} gives one way to understand the observed properties.  
According to this relation, we can estimate the intrinsic radio energy just by measuring the duration of the pulse. This can be a powerful tool for cosmology.  But the prerequisite is that we have to obtain a sufficient number of FRB samples \citep{Hashimoto2020b} so that we can make them standardizable in high accuracy.
In \citet[]{Hashimoto2019}, they used the publicly available catalogue, the FRBCAT by \citet[]{Petroff2016}, which compiled 68 verified FRB sample. However, they had to reduce the number of the sample down to 27 FRBs in order to secure a robust \cite[see][for details]{Hashimoto2019} sample for reliable analysis, which might cause possible selection bias.  
Even though they  noticed that there is a meaningful relation between the two important parameters, 
the number of samples they used to derive the relationship is much smaller than that of currently available FRBs.  We attempt to see if the relationship will be significantly changed or not, when we test all available FRBs recently detected by CHIME. Also, there is a possibility that latent repeaters could be included in their final sample that was supposed to be the purely non-repeating sources. This means that how to classify FRB populations correctly is critical to avoid the contamination from any types we do not intend to include.

The Canadian Hydrogen Intensity Mapping Experiment (CHIME) have recently release their first FRB catalogue along with various parameters \citep[]{CHIMEcat2021}. 
Because this catalogue contains a larger number (591 sub-bursts) of samples obtained in a consistent manner, totally free from a heterogeneous mixture of data from different telescopes, it provides a timely opportunity to test the empirical relation between the duration and radio energy of non-repeating FRBs  \citep["duration-energy relation", hereafter]{Hashimoto2019}.  

\citet[][]{Chen2021}  applied an unsupervised machine learning (ML) method on the CHIME catalogue to classify these bursts.   
This method helps us avoid the contamination from repeating FRBs because it can decompose different types and divide them into separate groups, even in complicated multiple parameter spaces \citep[see Figs 4 -- 8 in][]{Chen2021}.  This enables us to tests the relation using  the CHIME catalogue.

The purpose of this work is to examine the duration-energy relation using comparatively homogeneous FRB samples  obtained by CHIME  to take advantage of larger sample size   as well as a newly developed ML classification to avoid possible contamination by latent repeating populations. 
This paper is organised as follows.  Sec. \ref{Data}  briefly summarises the CHIME/FRB data and why we use a new classification by \citet{Chen2021}.   In sec. \ref{analysis}, we present the analysis and discussion using CHIME catalogue. In sec. \ref{Discuss}, we attempt to interpret the main results of this work.  In sec. \ref{summary}, we give conclusion and summary.

\section{Data and catalogue }
\label{Data}
 
We use the recent FRB catalogue released by Canadian Hydrogen Intensity Mapping Experiment (CHIME)  \citep[]{CHIMEcat2021}, which contains 536  FRBs (591 sub-bursts), at a frequency range between 400 and 800 MHz, which is an order of magnitude larger sample size compared to 
\citet{Hashimoto2019}.

To extract a certain burst type from the CHIME catalogue and deal with them more straightforwardly, we adopt the classification by \citet{Chen2021}. They used unsupervised machine learning (the Uniform Manifold Approximation and Projection; UMAP) method to train machine and showed the CHIME FRBs can be successfully classified into two categories, i.e., repeating and non-repeating bursts. Also, their results  showed that both repeating/non-repeating bursts can be divided further into a number of sub-groups: repeating bursts can be separated into three different sub-groups   while non-repeating sources are divided into six.  A fraction ($\sim$40$\%$) of one-off bursts in the CHIME catalogue  was reclassified as repeater candidates. These repeater candidates showed distributions different from what the rest of the one-off events showed in multiple parameter spaces tested in the UMAP method \citep{Chen2021}. 

\begin{figure}
    \includegraphics[width=\columnwidth]{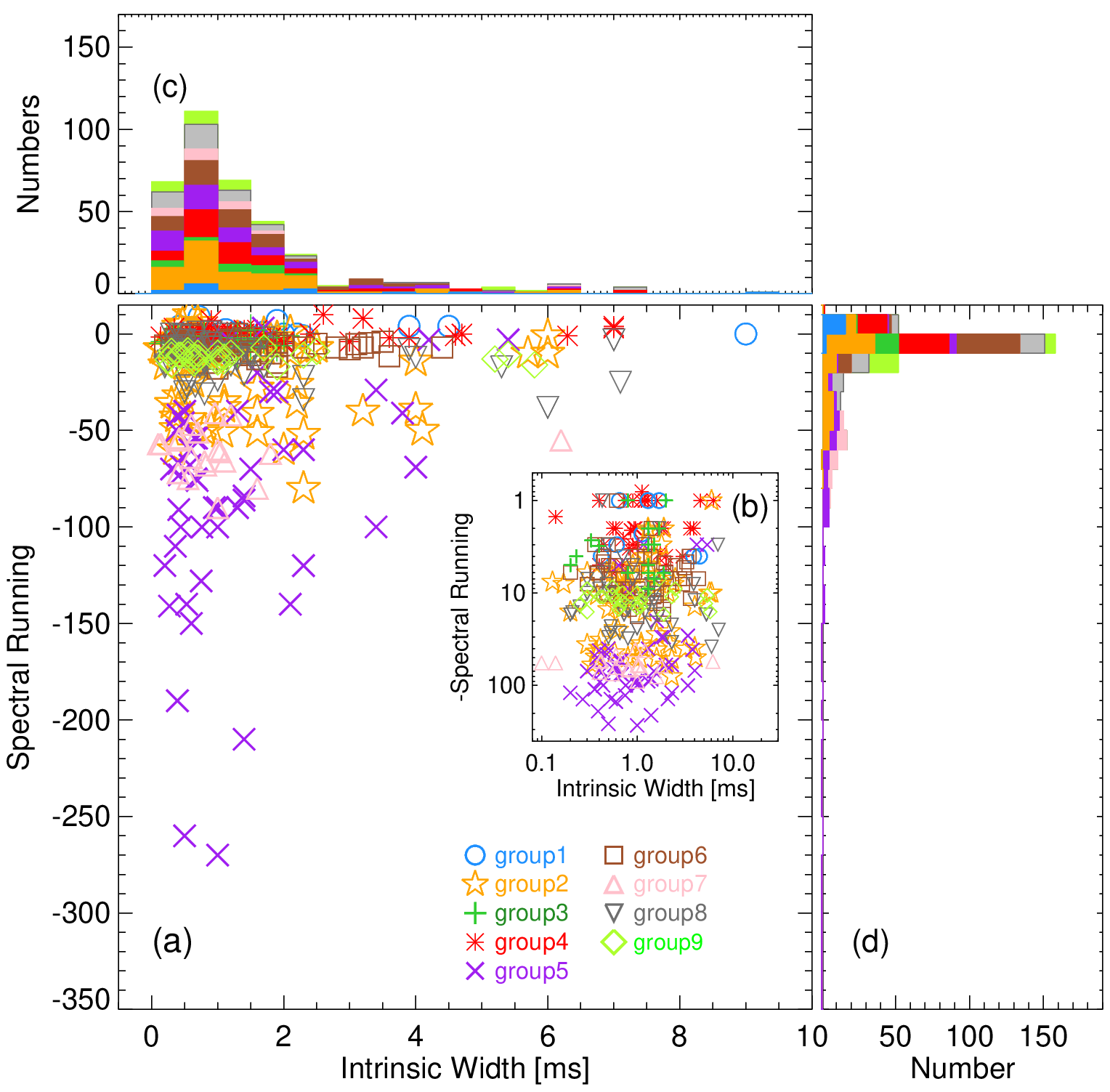}
    \caption{ Distribution of CHIME/FRB sources in the spectral running versus width plane (a). Different symbols indicate different types of sub-burst sources classified by unsupervised machine learning algorithm \citep[][nine sub-groups in this work]{Chen2021}.  In panel (b), it is given in logarithmic scale to see the crowded area (between 0 and -30 in spectral running) in more detail. Top (c) and right (d) panels show the histogram for each axis.
    }
    \label{spec_run}
\end{figure}

Machine learning methods could separate  different types  based on the model developed after training.  Therefore, we might be able to avoid the contamination from any types we do not want, i.e.,  both existing and/or latent repeater bursts. 
The detailed information about how the FRBs were classified using an unsupervised ML algorithm is described in \citet{Chen2021}. 
The details about how intrinsic duration and energy were derived and how the redshift ($z$) was calculated from the dispersion measure (DM) are described in   \citet{Hashimoto2019, Hashimoto2020c, Hashimoto2022}. We also give a summary in Appendix \ref{appendix}.

\section{Analysis and Results}
\label{analysis}

\subsection{Sub-groups of the CHIME non-repeaters}
 
We compare the energy and duration of the FRBs in the CHIME catalogue.
As shown by \citet{Pleunis2021}, the spectral running is one of the most efficient parameters showing the difference between one-off events and the repeater bursts (see Fig. 4 in \citealt[][]{Pleunis2021}). In their diagram showing the spectral running versus intrinsic width,  the repeater bursts are widely dispersed, while the one-off events show clustering compared to repeaters, forming a big major peak (in terms of probability) in the range of spectral running above $-25$ (as also shown by the histogram in Fig. \ref{spec_run}).  
But a small and broad bump spans over almost all the range of spectral running. Thus, if we cut at $-25$ in spectral running, we can easily divide them into two, and take the main peak to collect the one-off events sample, which looks statistically efficient.  
In the smaller bump, however, a faction of non-repeaters and repeaters are mixed together, being entangled complicatedly so that it is not straightforward to disentangle and separate them properly.  
Besides Fig. 4 in their paper,  \citet[][in sec.3]{Pleunis2021} gave a discussion on possible reasons for why different types (or different sub-groups by the UMAP, in Fig.  \ref{spec_run} of this work) are mixed in the same area.  A bias can be originated from the chromatic reduction in sensitivity when a burst is detected away from the beam centre. The relative sensitivity and beam response at the different positions can change the sub-structures and obscure the morphology of a burst, which might change the spectral shape and shift the position of the data point in Fig. \ref{spec_run} of this work or in Fig. 4 in \citet[][]{Pleunis2021}.  
However, these do not give explanations for how we can separate a selected group from this mixture so that we can filter out possibly hidden repeater types.
If we cannot extract what we want from this mixture, it will be extremely difficult to utilise all available non-repeaters, ruling out repeater bursts.

In the ML classification by \citet[]{Chen2021},  we can see that the different groups are occupying slightly different places with overlapping each other  as shown in Fig. \ref{spec_run} (a) and (b). 
As long as we use this classification, we can choose one or two specific groups or exclude them using the group ID, which enables us to perform assorted analysis  conveniently. Based on this classification applied to the CHIME/FRB catalogue,   we attempt to examine if there is something \citet{Hashimoto2019} did not notice owing to the insufficiency or heterogeneity of the sample as well as possible effects from hidden repeaters.    
\citet[]{Chen2021} used  thirteen parameters (highest frequency, peak frequency, scattering time, fluence, flux, spectral running, lowest frequency, radio energy, redshift, spectral index, width of sub-burst, boxcar width, and duration) from the CHIME catalogue to train machine and construct a model for classification.  Also, they checked how the UMAP model performed properly and how each parameter contributes to this unsupervised machine learning process, so that they presented those parameters in order of the feature importance based on their model  \citep[See Fig. 5 in][]{Chen2021}.

The first top six parameters (highest frequency, peak frequency, scattering time, fluence, and flux) turned out to be the most significant features in the ML process, when they were training the machine, which means their machine decided nine different sub-groups  based mostly on those major six parameters. The other parameters did not play significant roles in the classification compared to the top six parameters.   
Thereby we examined duration and energy (i.e., both parameters of duration-energy relation) as a function of the groups  classified by UMAP method as well as in terms of those top six parameters.  However, we did not find any significant trends or signatures: duration or energy does not have any prominent or particular dependencies on  sub-types once we select non-repeating FRBs.  
However, this is quite natural because both duration and energy did not play  important roles in the ML classification.  Therefore, duration-energy relation is not   closely related to a certain specific type of non-repeaters (although they are divided into subcategories).   Still, it is obvious that repeating FRBs have to be excluded  to correctly examine the duration-energy relation because they show totally different behaviour as shown by red triangles in Fig \ref{fig2}.

\subsection{Tests for redshift distribution}

Among the other parameters, redshift ($z$) has to be definitely examined, even though it was not a significant feature in the ML process. We have to check if there is any significant difference between different redshift bins or  if any evolutionary trend can be found. Here, it should be noted that  redshift was derived from the dispersion measure \citep[DM; see][] {Hashimoto2019, Hashimoto2020a, Hashimoto2020c}.   
In our analysis, we found that the tests of duration-energy relation with the CHIME/FRBs show 
redshift dependence: the correlation becomes significant  when we select  sample from $z=0$ towards the higher redshifts up to $z=0.3$, having the number of samples increased  gradually.

\begin{figure}
    \includegraphics[width=0.9\columnwidth]{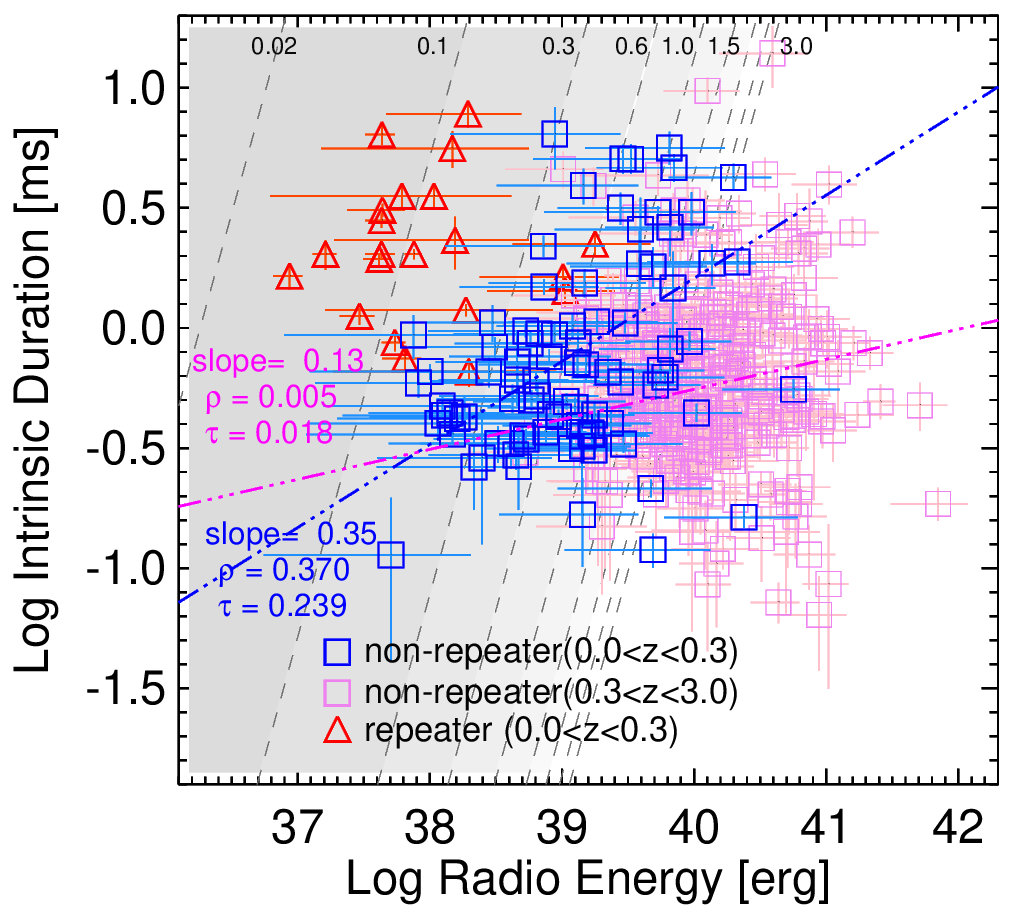}
    \caption{
    Distribution of CHIME/FRB sources in the duration vs energy plane.  Squares indicate non-repeating FRBs while triangles indicate repeating FRBs.  Non-repeaters (squares) are divided into two redshift bins (blue and magenta). Here the repeaters (red triangles) are presented only for the lower redshift bin ($z<0.3$) for comparison against non-repeaters at $z<0.3$ (blue squares). Dotted-dashed lines are the linear fits: the slope, Pearson's correlation coefficients ($\rho$), and Kendall's $\tau$  coefficient are given.  Only the blue boxes (local non-repeaters) show the positive correlation with a significant $\rho$ and $\tau$. Grey dashed lines show how the fluence threshold appears on this plane according to redshift (the numbers marked at the top)
    }
 \label{fig2}
\end{figure}

In Fig. \ref{fig2}, we show linear fits with the Pearson's ($\rho$) and  Kendall's $\tau$ correlation coefficient,  which are kinds of measures showing how strong a linear correlation between two data sets is.  If it is 0, there is no relation, while having 1 (or -1) means two data sets have a tight (or anti) correlation. 
The correlation becomes strongest at  0 $< z <$ 0.3, and when we include more non-repeater samples at higher redshift ($z > 0.3$), the correlation began to become less significant.  
Fig. \ref{fig2} summarises  overall aspects of current non-repeating bursts in the CHIME catalogue regarding this, dividing them into just two redshift bins by cutting at $z=0.3$. The distribution of non-repeaters (box symbols) is marked in blue and magenta, respectively. 
The non-repeaters at $z < 0.3$ (blue)  show the positive linear relation between the radio energy and duration, as reported by \citet[]{Hashimoto2019}.

\begin{figure}
    \includegraphics[width=0.9\columnwidth]{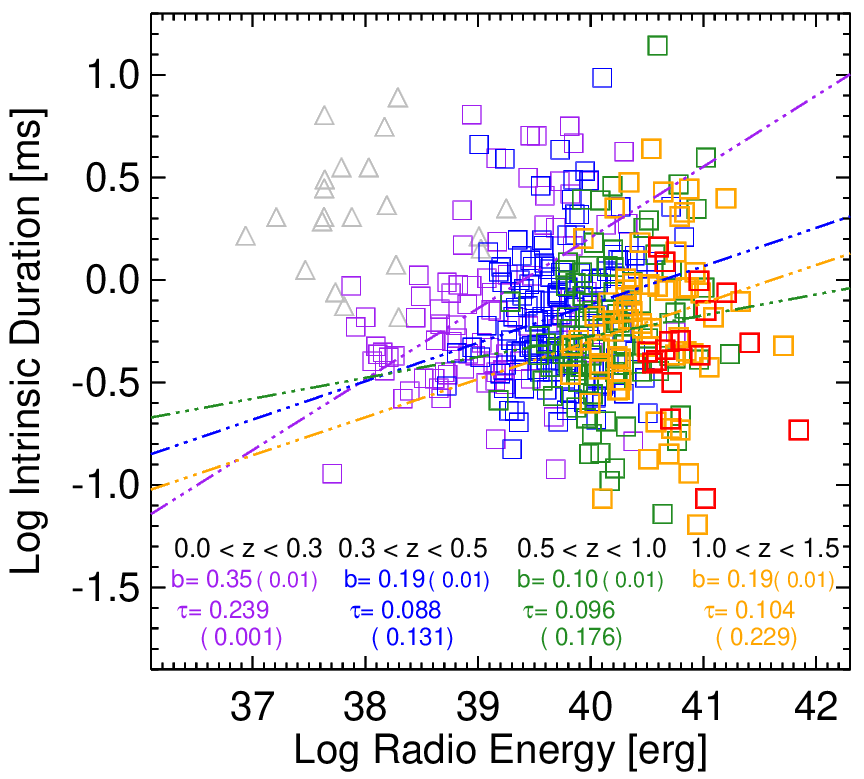}
    \caption{
    Distribution of non-repeating CHIME/FRBs   divided into four redshift bins as shown in purple, blue, green, and orange, respectively  for each redshift bin. (Error bars are omitted to avoid unnecessary crowdedness.) Red boxes represent the sources at $1.5 < z < 3.0$, but are not good enough to fit or derive a coefficient due to a small number. Straight lines are the linear fits to the non-repeaters at each redshift bin: the slope ($b$ with its 1-$\sigma$ error in the parentheses) and Kendall's $\tau$ coefficients. We present the significance at the bottom of this diagram in parentheses: the smaller number shows the higher significance.  The purple squares for the lowest redshift bin ($0.0<z<0.3$) correspond to the blue squares shown in Fig. \ref{fig2}.  
    }
 \label{fig3}
\end{figure}

The Pearson's correlation coefficient ($\rho$)  is 0.37 with the significance of   $0.021$ (i.e., lower than 0.05), which means it is unlikely to have occurred simply by chance.
But, the normal parametric method, Pearson’s correlation coefficient ($\rho$), might not be appropriate when we try to see the exact correlation because the data is affected by lower and upper limits (i.e., censored data).
Energy is derived from the fluence which is affected by the lower limit. Because of the uncertainty of localization and the significantly varying beam pattern and response, \citet[][]{CHIMEcat2021} assumed that each burst was detected along the meridian of the primary beam. Therefore, the fluences they provide are most appropriately interpreted as lower limits.  In this sense, we do not know the true fluence of individual events. If this effect can be taken into account (statistically), the true fluences will shift from the measured points by a certain amount on average towards higher fluences \citep[e.g.,][]{Macquart2018, Hashimoto2022}.
But this shift would not significantly affect the possible correlation between the duration and energy because the slope will basically stay the same. However, accurate localization is necessary to fully address this issue, which we leave as future works when the localization becomes available. 
Width ($w$) is derived from the \texttt{‘fitburst’} routine which is affected by the upper limit.  \citet[][]{CHIMEcat2021} used this routine, where the total-intensity data can be used to  measure values of $w$ larger than 100 $\mu$s. But for cases where the fitted value is smaller than this, they quote 100 $\mu$s as an upper limit.   Some of our data points  could be affected by this limit. But this is only for a small number of high-z samples: as shown in Fig. \ref{fig2}, there are four data points with duration shorter than 100 $\mu$s, which correspond to the green, orange and red data points in Fig. \ref{fig3} ($z>0.5$).  Therefore, the possible relation for the sample at $z<0.3$ would not be significantly affected by this limit. 
This is the reason why we present Kendall’s tau ($\tau$) rank correlation coefficient \citep[]{Feigelson2012.book}:  the  correlation coefficient for $z<0.3$ turned out to be 0.239 with the significance 0.001.  Therefore, we may be able to remark that the correlation appears weak currently, but statistically meaningful.

The FRB sources at $z > 0.3$ in Fig. \ref{fig2} do not show such a strong signal. This might be due to possible evolution according to redshift or  insufficient number of faint sample in higher-$z$.  However, we cannot make any concrete conclusion currently about this.   Here, we also show the repeaters at $z <$ 0.3 (red triangles),   for comparison against the non-repeaters at $z<0.3$ (blue squares) to describe why repeating types have to be excluded to test duration-energy relation correctly.
To obtain more significant and conclusive results at $z>0.3$, we need to collect more fainter but reliable sample to overcome the Malmquist bias at higher redshift. Future FRB surveys with accurate localization would significantly improve the fluence uncertainty. The accurate localization also allows measuring spectroscopic redshifts via the identification of their host galaxies. The spectroscopic redshifts will significantly reduce the uncertainty of energy calculation because most redshifts in our sample are derived from the dispersion measures.

In Fig. \ref{fig2},  we also show how the fluence threshold appears as a function of redshift. An observed fluence (Jy ms) can be transformed to restframe energy at a certain redshift. If there is no dependency on duration, this threshold will be described as a vertical line on the duration-energy plane. 
But, the detection limit of CHIME depends on the duration to the power of $1/2$ \citep{CHIMEFRB2019}. Due to this dependency, the threshold appears as a stright line with a slope in the duration-energy space. The fluence threshold may also vary according to the observing conditions of the bursts as well as various parameters such as the beam shape and spectral profiles.
\citet[][in Sec 2.3]{CHIMEcat2021}  recapitulated that the median across all bursts, at 95\% completeness, is approximately 5 Jy ms. 
Taking this value as a representative threshold, we can see how this threshold appears on the plane as shown by grey dashed lines along with redshifts (the numbers marked at the top) in Fig. \ref{fig2}.  

This threshold lines also represent 95\% completeness level at each redshift. For the FRB sample at $z<0.3$ (blue squares in Fig. \ref{fig2}),  90\% of them are brighter than the threshold line of $z=0.3$ (i.e., the completeness of them is at least higher than 95\%). The remaining 10\% are on the left side of this threshold line of $z=0.3$. But, if some of these 10\%  are at $z<0.1$, they are also higher than the 95\% completeness at $z=0.1$, as shown by the threshold line. Therefore, at least more than 90\% of our FRBs at $0<z<0.3$, which were used to derive the correlation coefficient, are higher than 95\% completeness.
Here, it could be worrisome that the threshold line has a slope, which could produce an artificial feature that looks like a correlation. 
\citet{Hashimoto2020a} performed the tests to see if the detection limit of the telescope mimics the duration-energy relation for non-repeating FRBs (Appendix C in their work). They generated artificial data, uniformly distributed in the duration-energy plane, having no ‘intrinsic’ correlation. They iterated the simulation 10,000 times using 120 artificial sources and concluded that the duration-dependent detection limit did not significantly mimic the correlation.

For more details,  Fig. \ref{fig3} shows the distribution of the non-repeating CHIME/FRBs at different redshift ranges, divided further into four different bins. The redshift distribution of the FRBs shows a trend along the energy  (i.e., along the horizontal axis).  Different groups represented by purple, blue, green and orange squares show a shift along the energy according to the redshift. Slightly overlapped with each other, they are moving from purple to orange,  which looks like a redshift evolution. However, we should keep it in mind that there is observation bias  because faint sources at higher-$z$ are more difficult to detect.   For each redshift bin, the slopes of the fitted lines change while all staying positive. The Kendall's $\tau$  correlation coefficient is getting insignificant at the higher redshift bins.

\begin{figure*}
    \includegraphics[width=0.9\textwidth]{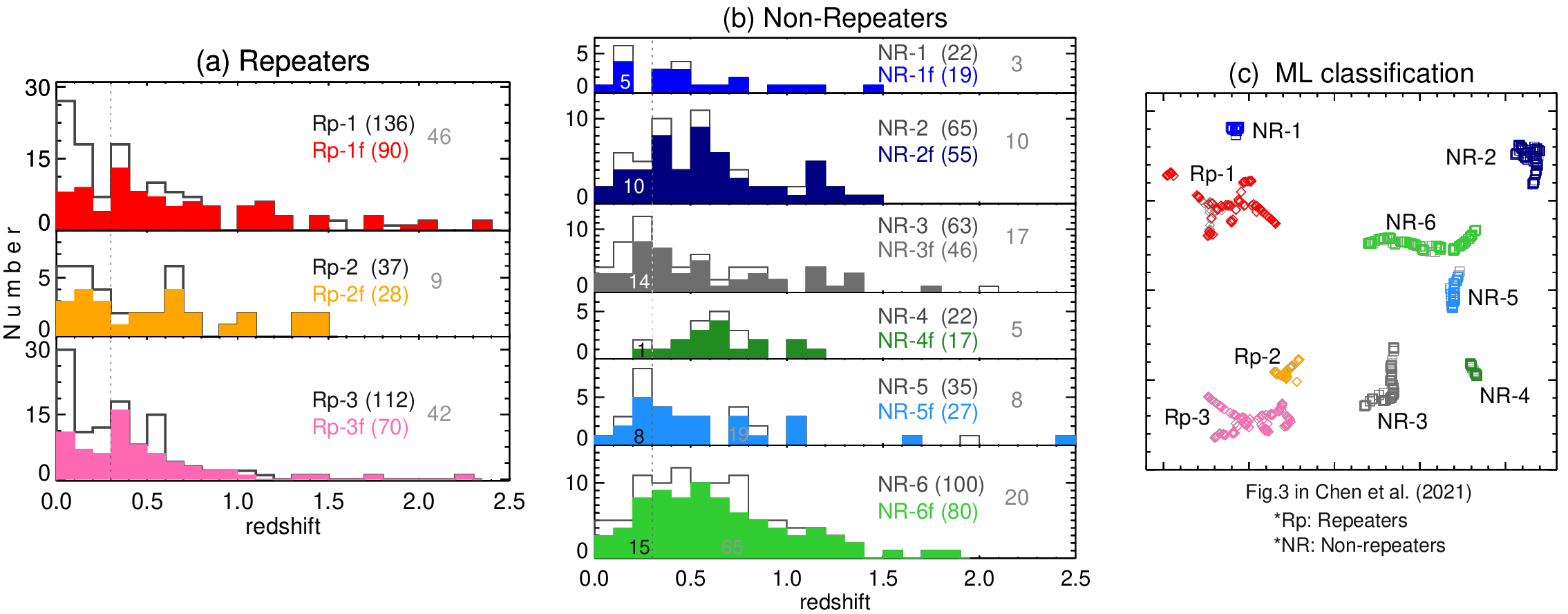}
    \caption{
    Source distribution as a function of redshift for both repeaters (a) and non-repeaters (b) classified by unsupervised machine learning (c).  Left: (a) three sub-groups of repeating types of CHIME/FRBs.      Middle: (b) six sub-groups of non-repeating types.  Right: (c) classification results of unsupervised machine learning, which shows the source distribution on a projected two-dimensional plane  describing how nine groups are located in term of relative distance to each other -- see Fig. 3 in \citet{Chen2021}. 
    The comparison before and after we   exclude flagged sample  is also given -- hollow histogram in grey lines shows the number of source before the exclusion.  The number of source is presented in the parenthesis.   The colour coding is the same in all three panels. 
    }
 \label{fig4}
\end{figure*}

The actual distribution of numbers as a function of redshift for each burst group classified by machine learning  (including both repeaters and non-repeater) is presented in Fig. 4.   Overall, most of the sub-groups (except for non-repeater group 4, deep green in the middle panel) show the peak  roughly between 0.3 - 0.5,  which tells us that the FRB events listed in the CHIME catalogue   have occurred in the rather local Universe ($z<1$). 
As \citet{Hashimoto2019} tried to use robust sample, we attempted to compare the sources with/without flagged FRB sample based on the information \texttt{(excluded\_flag, subb\_flag, and  subw\_upper\_flag)}  given in the catalogue \citep{CHIMEcat2021, Hashimoto2022}.

\begin{figure}
    \includegraphics[width=\columnwidth]{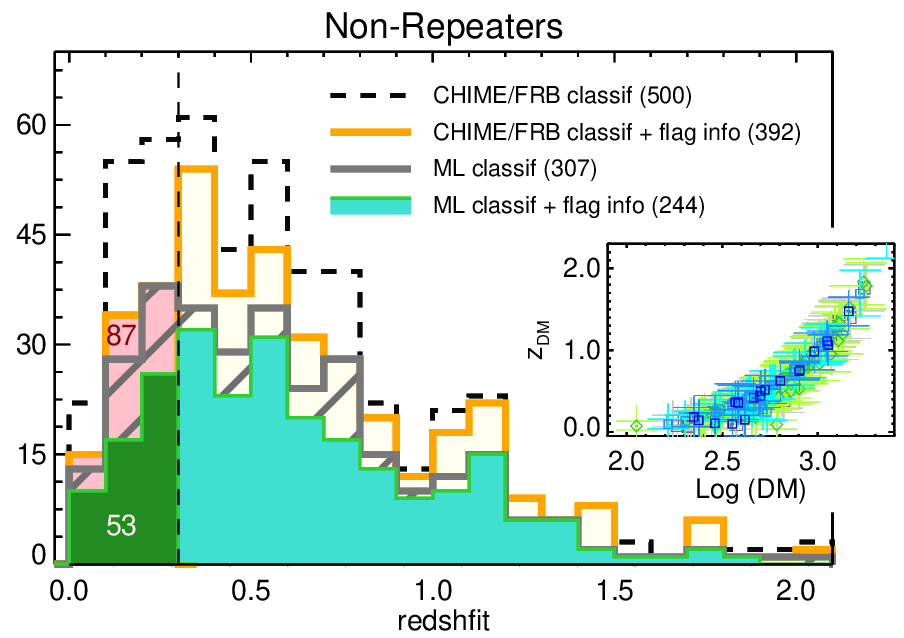}
    \caption{
    Overall distribution of non-repeating bursts regardless sub-group or type. Dotted histogram shows the distribution of one-off events originally classified by CHIME catalogue (500 sources). Orange line with yellow shaded region shows the one-off events originally classified by the CHIME but after excluding flagged sample (392 sources). There are only 87 sources at $z<0.3$ (a vertical line represents $z=0.3$).    Grey-line (dashed area) represents the distribution of one-off events classified by Machine Learning (307 sources). And then, we are left with 244 non-repeater sources when we apply flag information. There are 53 sources at $z<0.3$. An additional plot inside shows  the sizes of the errorbars for the redshift derived from the dispersion measure \citep{Hashimoto2019}.
    }
 \label{fig5}
\end{figure}

Excluding flagged samples reduces the actual number of available sources.
The grey hollow histograms and filled coloured histograms in  Fig. \ref{fig4} (a) and (b) show before  and after excluding flagged FRBs, respectively.
The number of sources for each sub-group is given in parenthesis. The number of excluded sources is presented in grey. The fact that the non-repeater group-4 (NR-4f, deep green in panel b) does not have  local sources at $z < 0.3$ indicates that  this group-4 is not a significant contributor to the duration-energy relation for $z<0.3$, even though we still do not have a clue about what they are (in term of the physical properties or origin). The rightmost panel (c)  shows a results from the source classification based on the unsupervised machine learning from  \citet [see Fig. 3 in the literature] {Chen2021}. This is a kind of  a projected two-dimensional plane  simplified from the complicated multi-dimensional hyperspace (e.g., a thirteen-parameter space of the ML).  Therefore, both axes are not expressed in the units that we can understand in any normal physical terms.  The distance to each data point is the only meaningful indicator where the same types are gathering close together, forming a cluster.  
The colour coding in all three panels is the same.  Therefore, the red data points or red histogram indicates the repeater (Rp) group-1, and the blue data points or blue histogram  represent the non-repeater (NR) group-1, in all three panels.

\begin{figure*}
    \includegraphics[width=0.70\textwidth]{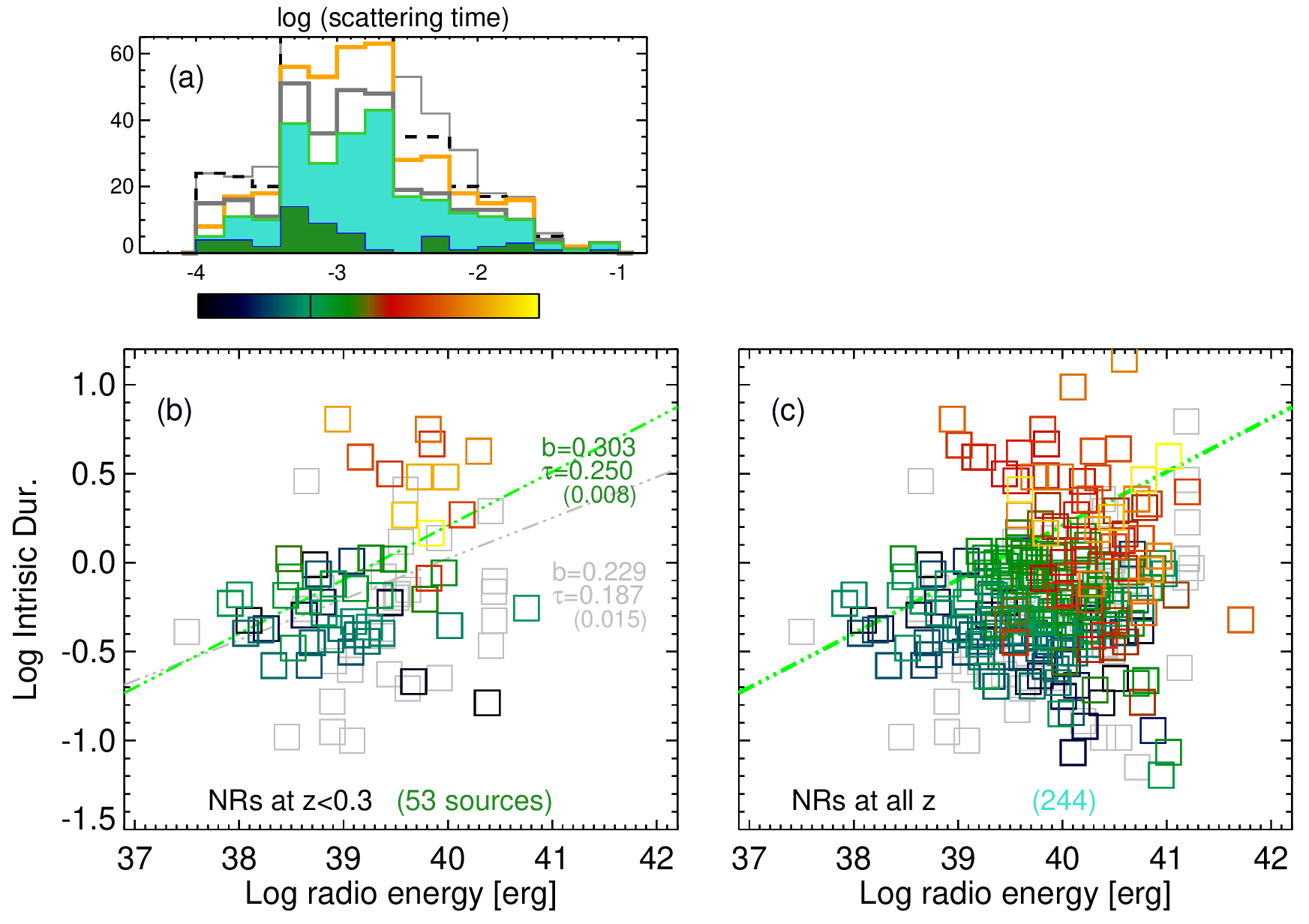}
    \caption{The distribution of non-repeater bursts in terms of scattering time (a) as well as on the duration-energy relationplane (b, c). The colours and line types of histograms (a) are the same as those in Fig. \ref{fig5}.  A colour bar under the histograms in panel (a), covering from -4 to -1.7 in scattering time, defines a colour coding used for the data points (squares) in panel (b) and (c).  In panel (b), the linear fit for the non-repeaters at $z<0.3$, comparing before and after we apply flag information, is given (a green and a grey lines). We have a stronger correlation ($\tau=0.250$) when we exclude flagged sample. Panel (c) shows the non-repeaters at all redshift range. A straight line is the green one from the panel (b).
    }
 \label{fig6}
\end{figure*}

For non-repeaters, in Fig. \ref{fig5}, we show the comparisons of CHIME classification versus ML classification as well as with/without applying flag information, as a function of redshift.    First, dashed hollow histogram shows the non-repeating sources originally classified by the CHIME catalogue. When we exclude the flagged sources, we have a smaller number of sample as shown by orange histogram (yellow shaded).  The pink area indicates the redshift range $z<0.3$, where 87 sources are included.
On the other hand, when we use ML classification, it becomes  different: because some ($\sim 40\%$) of original non-repeaters have been newly classified as repeater burst \citep{Chen2021} and excluded,  the number is already smaller, which is shown in grey thick line. In addition, if we use flag information, the distribution becomes the green filled histogram. This is approximately  half of the original non-repeater sample.  Eventually, if we cut at $z=0.3$, we have 53 sources (deep green shaded region in the histogram, Fig. \ref{fig5}). In an additional panel, we show the  redshift (derived from DM) versus DM. See sec. 2.2 in \citet{Hashimoto2022}  about the consistency with spectroscopic redshift and how uncertainties were propagated from DM -- the uncertainty of the redshift depends on the DM. Overall, the median size of the error-bars in this panel  is about $\pm$ $\sim$0.15.

\subsection{Scattering time  }

\begin{figure*}
    \includegraphics[width=\textwidth]{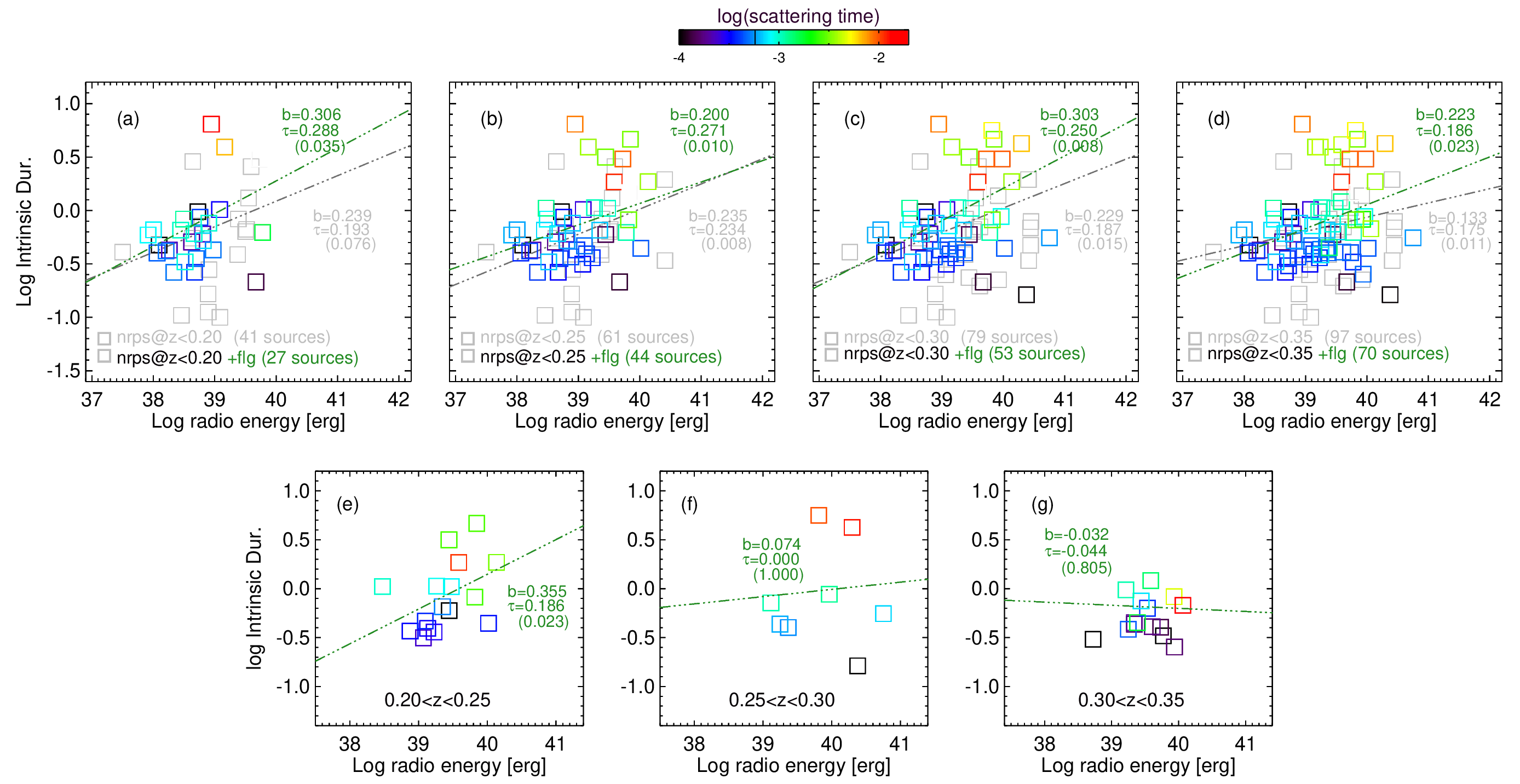}
    \caption{duration-energy relationtests in terms of the redshift range:  the distribution of the CHIME non-repeater sources and linear fits to them in slightly different redshift ranges (from a to d) at $z<0.35$.  Top panels show the comparison before and after applying flag information to exclude low-reliability sample. The number of sources are also given in parenthesis. Lower panels show the non-repeaters included in the finer redshift bins  between the upper panels.  Colour coding shows the scattering time as shown by a colourbar at the top. 
     }
 \label{fig7}
\end{figure*}

Scattering time is a measured quantity where propagation effects are contained. Scattering can diminish the detectability so it would be interesting to see if the scattering timescale shows any trend or characteristic with respect to the brightness of FRBs.
Scattering time, the third important parameter in ML \citep{Chen2021},  shows an interesting trend, as shown in Fig. \ref{fig6}. The scattering time vary/move diagonally along the line obtained by linear fit between energy and duration when the linear correlation is the strongest (i.e., $z<0.3$).   If we plot the histogram of non-repeaters as a function of `scattering time',  the distribution of sources (or histograms) in Fig. \ref{fig5} becomes the ones in Fig. \ref{fig6} (a).  The thick dashed line, orange line, grey line, and the green histogram  indicate the same as those  in Fig. \ref{fig5}, but the horizontal axis in Fig. \ref{fig6} (a) represents the logarithm of scattering time. 
A colour bar ranging from $-4$ to $-1.7$ under the histograms  is applied to the colour coding in Fig. \ref{fig6} (b) and (c).  In panel (b), green dot-dashed line is the linear fits (given with the slope and the Kendall's tau correlation coefficient). Grey data points and line show the results when we did not exclude flagged sample. To see the difference between the cases with/without the flagged sources, we carried out the derivation of the correlation coefficient for  both cases, as shown in Fig. \ref{fig6} and Fig. \ref{fig7}.  If we remove the flagged sample, we obtain a stronger correlation ($\tau=0.250$) even though it is not a significant change.  We can see how the scattering time varies  in the duration-energy plane.  Panel (b) shows the non-repeaters at $z<0.3$ while the panel (c) shows the distribution of all non-repeaters (with the green line from the panel b).   The scattering time shows a trend, which moves gradually from the lower left to the upper right, diagonally, along the linear fit.
Even though this doesn't give a direct information about duration-energy relation,   we can obviously see how scattering time changes, which might be a clue to approach to hidden physics behind the duration-energy relation (Revealing what this means, however, lies beyond the scope of this work).

In Fig. \ref{fig7}, we show more detailed  comparisons  in terms of the $z$-range.  The top panels, from(a) to (d), show the comparison between the different redshift ranges: from $z=0$ to $0.20, 0.25, 0.30$, and $0.35$, respectively.
In each panel, the green or grey  line indicates the linear fit to the data with/without applying the flags. The green line shows the cases when the flagged samples are excluded and always shows the better  (more significant $\tau$ correlation) coefficient  compared to grey line. Lower panels show the sources included only in the bin of the redshift increment between upper panels (while the upper panels show the source distribution in the accumulated ranges).  Thus, panel (e) shows the distribution of non-repeaters in the redshift range $0.20<z<0.25$, i.e., the increments between (a) and (b).  Panel (f) shows $0.25<z<0.30$ and panel (g) shows $0.30<z<0.35$, respectively.     
We can see  the strongest correlation coefficient when we cut at the redshift 0.25.  At the higher redshift ($z>0.25$), the correlation becomes weaker but stays positive up to $z=0.30$.   At the redshift $z>0.30$, the correlation becomes much weaker and no more positive. This is the reason why the correlation become weaker and less significant at $z>0.30$ (see also Fig. \ref{fig2} and Fig. \ref{fig3}).

\section{Discussion}
\label{Discuss}

In the above, we derive that the slope of duration-energy relation for non-repeating FRBs at $z<0.3$. Intriguingly, this slope value resembles that found for cosmological gamma-ray bursts (GRBs) $0.34\pm0.03$ by using a sample consisting of $386$ {\it Swift} GRBs with redshift measurements \citep{Tu2018}, which may indicate that the physical mechanism for these phenomena is similar. \citet{Tu2018} pointed out that such slope values close to $1/3$ could naturally arise from the magnetic reconnection, in which $E\propto B^2 l^3 \propto (\Delta t)^3$, where $B$ and $l$ are the magnetic field and the characteristic reconnection length, respectively\footnote{Here, the Alfv\'en velocity and the Mach number are both assumed to be constant and thus omitted for simplicity.}. If this is the case for FRBs, our result may support any theories that invoke magnetic reconnection as a mechanism for the energy release in FRBs. 
Still, a more accurate determination with a larger number of sample is needed to give better constraint on the emission mechanisms and progenitor models.

To utilise this relation in cosmology, it has to be well established first. If the relation would be confirmed in the future by a larger sample (with high-quality data at higher-$z$), the non-repeating FRBs could be used as an alternative distance measure (e.g., luminosity distance). The established relation would directly give the intrinsic energy from the duration, and we can construct the Hubble diagram showing the distance modulus (i.e., the difference between observed and absolute magnitude) versus luminosity distance (or redshift derived from the DM-$z$ relation). 

However, the current relation in our work does not appear to be clear/strong since the distribution of data points (e.g., $z<0.3$) in the duration-energy plane shows a scatter with large error bars. This wide distribution of the data points is mainly originated from the given parameters, how we derived redshift and energy for FRBs, and how we divided the FRB types and redshift bins. 
The dispersion measure (DM, representing free-electron column density) is an important parameter, which is used to derive the redshift and energy. DM is composed of a few components from different origins: from us to the FRB host, interstellar medium in our Milky Way Galaxy (DM$_{\rm MW}$) and hot gas in its halo (DM$_{\rm halo}$), intergalactic medium (DM$_{\rm IGM}$), and the FRB host galaxy (DM$_{\rm host}$), which is expressed as,  DM$_{\rm obs}$ = $ \rm{ DM_{MW} + DM_{halo} + DM_{IGM} + DM_{host} }$. The measurement error in observed DM (i.e., DM$_{\rm obs}$) does not play a significant role because it can be measured with a precision of less than 1 $\rm pc/cm^{3}$. 
In terms of the redshift uncertainty, the line-of-sight fluctuation of DM$_{\rm IGM}$ and DM$_{\rm host}$ uncertainty could be dominant. The uncertainty of DM$_{\rm host}$ can be as high as a few hundreds of $\rm pc/cm^{3}$. However, assuming an average value of DM$_{\rm host}$ $=50/(1+z)$ $\rm pc/cm^{3}$, \citet{Hashimoto2022} confirmed that DM-derived redshifts are consistent with spectroscopic redshifts within the line-of-sight fluctuation of DM$_{\rm IGM}$ for the CHIME and FRBCAT samples.

In the unsupervised UMAP classification, the existence of 9 different groups does not necessarily mean that there are 9 different physical mechanisms that generate FRB phenomena.  Since the input parameters used in the ML classification are from the CHIME FRB catalogue, we use the ML results only to avoid the effects from the possible hidden repeater types.   The main key in the ML method is that there are some differences (between the two populations) which might have been projected to the CHIME/FRB parameters.  It should be also noted that inevitable uncertainties and biases are contained, e.g., originated from the small sample of repeating sources and the imbalance between the number of apparently non-repeating and the repeating sources, the use of multiple bursts from the same repeating sources.  As discussed by \citet[][]{CHIMEcat2021} and \citet[][]{Pleunis2021}, the data could be biased because of the instrumental effects, beam shape, sky coverage, sensitivity threshold, etc.  For example, the submillisecond structure of bursts could be unresolved and a multicomponent might be interpreted as having just one component.  As long as we use duration given in the catalogue based on ‘fitburts’,  the relation between energy and duration at $z<0.3$ is not significantly affected by the upper limit, 100$\mu$s. Here, to derive a relation reasonably, we obtained Kendall’s tau coefficient, which takes the censoring and truncation into account.

Using permutation feature importance, a technique for model inspection, \citet[][]{Chen2021} found out that the highest frequency is the most important parameter and the second important one is the peak frequency.  Attempting to obtain a physically motivated interpretation of why these frequencies are most important, they presented a speculative remark that the spectral shape is probably a key factor, which is projected to the CHIME frequency range.

\section{Summary and Conclusion}
\label{summary}
In this work, we attempted to examine the duration-energy relation  \citep[][]{Hashimoto2019} based on a much larger sample size  and carried out further analysis to understand the properties of the  relation  in terms of the FRB parameters in the new CHIME catalogue.
We found a meaningful correlation   at the lower redshift range ($z<0.3$).   At higher redshift bins ($z>0.3$), the relation becomes less significant and can not reproduce the   relation reported by \citet[]{Hashimoto2019}. 
This is possibly due to the insufficient (especially, lack of faint) FRBs at high-$z$, and/or possible evolution of the relation. 
A larger number of fainter FRBs at high-$z$ are awaited to conclude on the relation at high-$z$.

We summarize the duration-energy relationtests in this work as follows.

(1) We used larger sample compared to  \citet[]{Hashimoto2019}: from the CHIME/FRB catalogue, twice the sample size (e.g., 53 non-repeating sample at $z<0.3$)   contributes to the positive correlation. 

(2) duration-energy relation may not be originated from any sub-groups/types of non-repeating FRBs (it might be an integrated characteristic of all non-repeater FRBs).

(3) The redshift range lower than 0.30 is  important for duration-energy relation: the main driver of the duration-energy relation is the non-repeating FRBs at the redshift $z<0.25$.

(4) How to separate/exclude the repeating bursts from the mixture of all different types of FRBs in complicated parameter spaces is essential to examine the behaviours of non-repeating FRBs. 

(5) The slope between the duration and energy relation (at $z<0.3$) resembles that found for cosmological GRBs (0.34 ±0.03), which may imply that the physical mechanism could be related to the magnetic reconnection.
 
(6) Scattering time shows an increasing trend along the straight line representing a linear fit between radio energy and duration.

\section*{Acknowledgements}
We would like to thank the anonymous reviewer for meticulous reading and constructive comments that helped improve the quality and clarity of this paper. TG and TH acknowledge the supports of the Ministry of Science and Technology (MoST) of Taiwan through grants 108-2628-M-007-004-MY3 and 110-2112-M-005-013-MY3, respectively.  This work used high-performance computing facilities operated by the Centre for Informatics and Computation in Astronomy (CICA) at National Tsing Hua University (NTHU).  This equipment was funded by the Ministry of Education of Taiwan, the Ministry of Science and Technology of Taiwan, and NTHU.
This research has used NASA's Astrophysics Data System.

\section*{Data Availability}
 
The data underlying this paper (the CHIME FRB data) is available at \url{https://www.chime-frb.ca/catalog}.  
The FRB classification based on the unsupervised machine learning is available in the online supplementary material by \citet[]{Chen2021}.
The main parameters (e.g., redshift, duration, radio energy, etc., and their uncertainties) are available at  the online supplementary material of \citet{Hashimoto2022}: \url{https://academic.oup.com/mnras/article/511/2/1961/6507586\#supplementary-data}.


\bibliographystyle{mnras}
\bibliography{E_w_CHIME_FRBs}

\appendix
\section{How the physical parameters were derived}
\label{appendix}

\subsection{Redshift}
\label{calc_redshift}
We follow the same manner as that of \citet{Hashimoto2020c, Hashimoto2022} to derive the redshift of each FRBs except for the ones whose spectroscopic redshifts were decided. As mentioned in sec. \ref{Discuss},
\begin{equation}
\label{eqDM}
{\rm DM}_{\rm obs}={\rm DM_{\rm MW}}+{\rm DM}_{\rm halo}+{\rm DM}_{\rm IGM}+{\rm DM}_{\rm host}.
\end{equation}

We adopt DM$_{\rm MW}$ modelled by \citet{Yao2017}, DM$_{\rm halo}=65$ pc cm$^{-3}$ (\citealt{Prochaska2019}, see also \citealt{Yamasaki2020} for direction-dependent model), and DM$_{\rm host}=50.0/(1+z)$ pc cm$^{-3}$ following \citet{Macquart2020}. 
The DM$_{\rm IGM}$ averaged over the line-of-sight fluctuation is described as a function of redshift  with some assumptions on the cosmological parameters \citep[e.g. Equation 2 in][]{Zhou2014}, i.e., for a flat Universe,
{
\begin{equation}
\label{eqDMIGM}
\begin{split}
&\mathrm{DM}_{\mathrm{IGM}}(z)=\Omega_{\mathrm{b}} \frac{3 H_{0} c}{8 \pi G m_{\mathrm{p}}} \times \\
&\int_{0}^{z} \frac{\left(1+z^{\prime}\right) f_{\mathrm{IGM}}\left(z^{\prime}\right)\left(Y_{\mathrm{H}} X_{\mathrm{e}, \mathrm{H}}\left(z^{\prime}\right)+\frac{1}{2} Y_{\mathrm{p}} X_{\mathrm{e}, \mathrm{He}}\left(z^{\prime}\right)\right)}{\left\{\Omega_{\mathrm{m}}\left(1+z^{\prime}\right)^{3}+\Omega_{\Lambda}\left(1+z^{\prime}\right)^{3\left[1+w\left(z^{\prime}\right)\right]}\right\}^{1 / 2}} d z^{\prime}.
\end{split}
\end{equation}
}
 Here $X_{\mathrm{e}, \mathrm{H}}$ and $X_{\mathrm{e}, \mathrm{He}}$ are the ionisation fractions of the intergalactic hydrogen and helium, respectively.  $Y_{\mathrm{H}}=\frac{3}{4}$ and $Y_{\mathrm{p}}=\frac{1}{4}$ are the mass fractions of H and He.   $f_{\mathrm{IGM}}$ is the fraction of baryons in the IGM.  The equation of state of dark energy is expressed as $w$.   We assumed $X_{\mathrm{e}, \mathrm{H}}=1$ and $X_{\mathrm{e}, \mathrm{He}}=1$, which are reasonable up to $z \sim 3$ because the IGM is almost fully ionised.
We adopted $f_{\mathrm{IGM}}$ = 0.9 at $z>1.5$ and $f_{\mathrm{IGM}}= 0.053z+0.82$ at $z\leq1.5$ following literature \citep[]{Zhou2014}.  Taking DM$_{\rm host}$ and \ref{eqDMIGM}, the right term of \ref{eqDM} is a function of redshift. To estimate the redshift uncertainty of each FRB, we performed Monte Carlo (MC) simulations, where randomised errors are added to DM$_{\rm obs}$ and DM$_{\rm IGM}$. 
The randomised errors follow Gaussian probability distributions with standard deviations of $\delta$DM$_{\rm obs}$ and $\sigma_{\rm DM_{\rm obs}}$.
We conservatively assume the highest $\sigma_{\rm DM_{\rm obs}}$ estimated from cosmological simulations of structure formation \citep{Zhu2018}.
Since $\sigma_{\rm DM_{\rm obs}}$ is estimated as a function of redshift up to $z=2$ \citep{Zhu2018}, we linearly extrapolate $\sigma_{\rm DM_{\rm obs}}$ towards higher redshifts \citep[see][for details]{Hashimoto2020c}.

\subsection{Energy integrated over the rest-frame 400 MHz width}
\citet{Hashimoto2020a} and \citet{Hashimoto2020c} utilise the time-integrated luminosity in units of erg Hz$^{-1}$, which is calculated from the observed fluence.
In this work, we use the energy in units of ergs, i.e., fluence integrated over the frequency, as an indicator of the brightness of FRBs based on the following reasons.
The first reason is that some FRBs show complicated sub-structures in their light curves.
Such shapes of light-curve and peak flux density would highly depend on the time resolutions of instruments. 
The fluence is less affected by the finite time resolution of instruments \citep[e.g.][]{Macquart2018}, which allows us to mitigate systematic differences when comparing with FRBs detected with other telescopes \citep[e.g.][]{Hashimoto2020a,Hashimoto2020c}.
The second reason is that FRBs detected with CHIME also show complicated spectral shapes \citep{Pleunis2021}.
\citet{Pleunis2021} presented the diverse spectral shapes: non-repeating FRBs tend to show broad-band power-law like shapes whereas repeating FRBs tend to show narrow-band Gaussian-like spectral shapes.
The $k$-correction for such diverse spectral shapes would be highly uncertain since complicated extrapolations of the spectral shapes are necessary.
To minimise such uncertainty, we integrate the fluence over the frequency to calculate observed energy ($E_{\rm obs}$) for each FRB. 
This frequency integration is described as $E_{\rm obs}=$ fluence $\times\left(\frac{400\times10^{6}}{{\rm Hz}}\right)$ because the fluence in the first CHIME/FRB catalogue is the band-averaged value over the CHIME frequency width of 400 MHz \citep{CHIMEcat2021}.

The observed frequency width of 400 MHz corresponds to different frequency widths at the rest-frame depending on the redshifts of FRBs.
For a fair comparison at different redshifts, we use the integration over 400 MHz widths at the rest-frame.
We define the integration width in the observer-frame, $\Delta\nu_{\rm obs,itg}$, which corresponds to the 400 MHz at the rest-frame, i.e., $\Delta\nu_{\rm obs,itg}=400/(1+z)$ MHz.
The observed energy integrated over the rest-frame 400 MHz width is $E_{\rm obs,400}  =  F_{\nu}\left(\frac{400\times10^{6}}{{\rm Hz}}\right)$, if $\Delta\nu_{\rm obs,itg}\geq \Delta\nu_{\rm obs,FRB}$, and  $F_{\nu} \left(\frac{400\times10^{6}}{{\rm Hz}}\right)\left(\frac{\Delta\nu_{\rm obs,itg}}{\Delta\nu_{\rm obs,FRB}}\right)$, if $\Delta\nu_{\rm obs,itg}<\Delta\nu_{\rm obs,FRB}$,
where $F_{\nu}$ is the observed fluence and $\Delta\nu_{\rm obs,FRB}$ is the frequency width in which FRB is detected.
For each FRB without multiple sub-bursts, $\Delta\nu_{\rm obs,FRB}$ is the difference between highest frequency and lowest frequency in the first CHIME/FRB catalogue. 
For each FRB with multiple sub-bursts, the maximum (minimum) value of highest (lowest) frequency is adopted to calculate $\Delta\nu_{\rm obs,FRB}$ because there is no frequency gap between the sub-bursts in the catalogue.
The ratio, $\Delta\nu_{\rm obs,itg}/\Delta\nu_{\rm obs,FRB}$, approximately takes the overflowed energy out of the rest-frame 400 MHz width into account.

Following \citet{Macquart2018b}, we calculate the rest-frame isotropic radio energy ($E_{\rm rest,400}$) for each FRB.
By integrating Eq. 8 in \citet{Macquart2018b} over the frequency, $E_{\rm rest,400}$ is described as
\begin{equation}
\label{eq:Erest400}
E_{\rm rest,400}=4\pi d_{l}^{2}E_{\rm obs,400}/(1+z),
\end{equation}
where $d_{l}$ is the luminosity distance to the redshift of FRB.
The uncertainty of $E_{\rm rest,400}$ ($\delta E_{\rm rest,400}$) includes the error propagation of $\delta$DM$_{\rm obs}$, $\sigma_{\rm DM_{\rm obs}}$, and the uncertainty of $F_{\nu}$ ($\delta F_{\nu}$) through  \ref{eqDM}  to \ref{eq:Erest400}.
To estimate $\delta E_{\rm rest,400}$, we performed the same manner as the MC simulations for the redshift uncertainty (see Section \ref{calc_redshift}) with 10,000 iterations.

  \bsp typesetting comment
\label{lastpage}
\end{document}